\documentclass[aps, amsfonts, amsmath,nofootinbib, showpacs, byrevtex, preprintnumbers, showkeys, preprint, prd,natbib]{revtex4-1}   
\usepackage{graphicx}
\usepackage{color}
\usepackage{subfigure}
\usepackage{bbm}

\newcommand{\cD}{{\cal D}}

\newcommand{\bi}{\bigskip}

 \newcommand{\no}{\noindent}  
\newcommand{\bea}{\begin{eqnarray}}
\newcommand{\eea}{\end{eqnarray}}
\newcommand{\be}{\begin{equation}}
\newcommand{\ee}{\end{equation}}

\newcommand*{\varpm}{\mathbin{\ooalign{\hfil$\pm$\hfil\cr\hfil\raise-.3ex\hbox{$\scriptscriptstyle(\mkern16mu)$}\hfil}}}

\newcommand{\sli}{\sum\limits}

\newcommand{\il}{\int\limits}

\newcommand{\ZZ}{\mathbbm{Z}}

\newcommand{\Id}{ \mathbbm{1}}

\usepackage[dvipsnames]{xcolor}
\usepackage{soul}
\usepackage{changes}
\definechangesauthor[name={Hugo Reinhardt}, color=Green]{hr}
\definechangesauthor[name={Luis Oxman}, color=Red]{ox} \setremarkmarkup{[\textsc{#2}]}

\begin{document}
\setstcolor{Green}

\title{Effective theory of the D = 3 center vortex ensemble}

\author{L.~E.~Oxman}
\affiliation{Universidade Federal Fluminense\\
	Instituto de F\'{i}sica\\
	Campus da Praia Vermelha, 24210-340, Niter\'oi, RJ\\
	Brazil}
\author{H.~Reinhardt}
\affiliation{Institut f\"ur Theoretische Physik\\
Auf der Morgenstelle 14\\
D-72076 T\"ubingen\\
Germany}
\date{\today}
%


\begin{abstract}
By means of lattice calculations, center vortices have been established as the infrared dominant gauge field configurations of Yang-Mills theory. In this work, we investigate an ensemble of center vortices in D = 3 Euclidean space-time dimension where they form closed flux loops. To account for the properties of center vortices detected on the lattice, they are equipped with tension,  stiffness and a repulsive contact interaction. The ensemble of oriented center vortices is then mapped onto an effective theory of a complex scalar field with a U(1) symmetry.
For a  positive tension, small vortex loops are favoured and the Wilson loop displays a perimeter law   while for a negative tension, large loops dominate the ensemble. In this case the U(1) symmetry of the effective scalar field theory is spontaneously broken and the Wilson loop shows an area law. To account for the large quantum fluctuations of the corresponding Goldstone modes, we use a lattice representation, which results in an XY model with frustration, for which we also study the Villain approximation. 
\end{abstract}
\maketitle
Pfad: /paper/unpublished/paper-efftheory/eff-theory.tex
\bi

\no
\section{Introduction}
\bi

\no
The essential features of the QCD vacuum are confinement and the spontaneous breaking of chiral symmetry. A thorough understanding 
of these two phenomena, and of the infrared sector of QCD in general, is still lacking. However, substantial progress has been 
achieved during the last twenty years in identifying the relevant infrared degrees of freedom of QCD (or at least of Yang-Mills theory). 
From these studies consistent confinement pictures have emerged: The center vortex picture \cite{tHooft:1977nqb,*Vinciarelli:1978kp,*Yoneya:1978dt,*Cornwall:1979hz,*Mack:1978rq,*Nielsen:1979xu}, the dual Meissner effect \cite{tHooft:1981bkw,*Mandelstam:1974pi,*Nambu:1974zg,*Baker:1991bc} and the Gribov-Zwanziger 
picture \cite{Gribov:1977wm,*Zwanziger:1988jt}. These different scenarios do not contradict each other but turn out to be just different manifestations of the same phenomena in different
gauges. Center vortices detected on the lattice by the method of center projection \cite{Deb+97} show the proper scaling behaviour towards 
the continuum limit \cite{Langfeld:1997jx,DelDebbio:1998luz} only in the so-called maximal center  gauge. Analogously, magnetic monopoles are detected after Abelian 
projection and show proper scaling only in the maximal Abelian gauge \cite{Kronfeld:1987ri,*Suzuki:1989gp,*Chernodub:1994pw,*Bali:1996dm}. (The so-called ``indirect maximal center gauge'' is done on top 
of the ma\-xi\-mal Abelian gauge). Finally, the Gribov-Zwanziger picture \cite{Gribov:1977wm,*Zwanziger:1988jt} has been established in Coulomb gauge \cite{Feuchter:2004mk,*Reinhardt:2004mm,*Epple:2006hv}.  
\bi
 
\no 
Magnetic monopoles are attached to center vortices \cite{Ambjorn:1999ym} and change the direction of the flux of center vortices \cite{Reinhardt:2001kf}\footnote{The orientation of the flux of center vortices is irrelevant for their confining properties but crucial for their topological charge \cite{Reinhardt:2001kf} and for spontaneous breaking of chiral symmetry.}.
Therefore, condensation of center vortices in the confined phase implies also the condensation of magnetic monopoles and thus the  
dual Meissner effect. Center vortices as well as magnetic monopoles live on the Gribov horizon in both Coulomb and Landau gauge \cite{GreOleZwa04}. 
Configurations on the Gribov horizon give rise to an infrared diverging ghost form factor, a necessary condition for the 
Gribov-Zwanziger picture to be realized. When center vortices are eliminated from the ensemble of gauge field configurations contributing 
to the Yang-Mills functional integral, the ghost form factor becomes infrared finite and confinement is lost \cite{Burgio:2015hsa}. Furthermore, the 
Gribov-Zwanziger mechanism in Coulomb gauge presupposes the dual Meissner effect \cite{Reinhardt:2008ek}. 
\bi

\no
Center vortices are localized field configurations for which the Wilson loop operator becomes a center element of the gauge group
if the latter is non-trivially linked to the center vortex, see eq. (\ref{cele}) below. Lattice calculations performed in the maximal
center gauge have provided mounting evidence that the center vortices are the dominant infrared degrees of freedom of Yang-Mills
theory: When center vortices are removed from the gauge field ensemble of the lattice functional integral \cite{deForcrand:1999our} the string tension 
is lost, chiral symmetry is restored \cite{Gattnar:2004gx} and topological charge is lost \cite{BerEngFab01}. The emergence of the string tension can be easily understood in 
a random center vortex model \cite{Engelhardt:1999fd}. Furthermore, center vortices give also a simple explanation of the deconfinement phase 
transition \cite{Engelhardt:1999fd}.
\bi

\no
In an idealized picture, which is realized, in particular, after center projection on the lattice, center vortices are point-like 
objects in D = 2, closed strings in D = 3 and closed surfaces in D = 4. They are closed by the Bianchi identity and live on the 
dual lattice. The gross features of center projected Yang-Mills theory, like the emergence of the string tension or the deconfinement 
phase transition, can be reproduced in a center vortex model with an action given by the vortex area plus a penalty for the curvature 
of the vortices \cite{EngRei00a,*Engelhardt:2003wm,*Quandt:2004gy}. The latter accounts for the stiffness of the vortices. In D = 4 such a model has to be numerically simulated on the lattice
since a (continuum) string theory cannot be treated analytically. Since D = 3 Yang-Mills theory has the same infrared properties as 
D = 4 it is useful to investigate the center vortex model in D = 3, where vortices are closed loops.

In this paper we  study the 
ensemble of closed center vortices in D = 3 exploiting the fact that the partition function of a gas of one dimensional objects can be represented 
by a complex scalar quantum field theory \cite{Stone:1978mx,*Samuel:1977vy,*Samuel:1979mq}. 
 Within this theory, we calculate the Wilson loop. To keep the soft Goldstone modes of the complex scalar field, we resort to a lattice representation which leads to the $3D$ XY model with frustration for which we also study the Villain approximation.  Since we are interested here in the Wilson loop, we consider only oriented center vortices as the orientation (of the flux) of center vortices is irrelevant for the confining properties. Non-oriented center vortices arise in the presence of magnetic monopoles \cite{Reinhardt:2001kf}. Such vortices were considered in $D = 3$ in ref. \cite{deLemos:2011ww}. Let us also mention that an ensemble of closed center vortices in $3D$, generated by Monte Carlo methods applied to lines with stiffness that can grow, shrink and reconnect, 
was recently considered in Ref. \cite{Altarawneh:2016ped}. Furthermore, Monte Carlo simulations to 
explore the statistical properties of the $3D$ XY model using a disorder parameter that creates flux vorticity were carried out in Ref. \cite{DiCecio:1996ty}.  

\bi

\no
The organization of the paper is as follows: In the next section we consider the partition function of an ensemble of ideal center vortices in D=3 Euclidean space time dimensions in the presence of a Wilson loop. The center vortex loops are endowed with tension, stiffness and a binary repulsive interaction. The latter is linearized by means of a real scalar field. Then the partition function of the vortex ensemble is reduced to the quantum transition amplitude of a single center vortex in an „external“ scalar field. In section \ref{sectIII} this amplitude is expanded in leading order in spherical harmonics and transformed into an effective theory of a complex scalar field. This field develops a non-trivial vacuum expectation value. To study the quantum fluctuations of this field, in section \ref{sectIV} the theory is translated to a lattice where it results in an XY model, which is also investigated in the Villain approximation. Finally the Wilson loop is calculated at low and high temperatures in section \ref{sectV}. A short summary and our conclusions are given in section \ref{sectVI}.

\section{Ensemble of thin center vortices in $3$D}
\label{ensemble-cv}

Lattice calculations in pure $SU(N)$ Yang-Mills (YM) theory have established center vortices as the essential degrees of freedom underlying confinement. While center vortex removal leads to a perimeter law, the projection to the center vortex ensemble gives an area law, with the $N$-ality properties observed in the complete Monte Carlo simulations. 
By the Bianchi identity, center vortices form closed manifolds, i.e. closed loops in D = 3, of color electric or magnetic flux. 
\bi

\no
The effect of a thin center vortex on a Wilson loop $ \langle W_{\cal C} \rangle $ is topological. In a three dimensional Euclidean spacetime, 
when a closed center vortex worldline $l$ links the Wilson loop ${\cal C}$ the latter gains a factor
\begin{equation}
\mathfrak{z} ({\cal C}) =  z^{L (l,{\cal C})} \;,
\label{cele}
\end{equation}
where $z$ is an element of the center $Z (N)$ of the gauge group $SU(N)$ and $L (l, C) \in \ZZ$ is the linking number, which is a topological invariant that counts the number of times the loop $l$ winds around the loop ${\cal C}$, with a sign that depends on whether $l$ has positive or negative orientation with respect to ${\cal C}$. An explicit integral formula was given by Gauss,
\begin{equation}
\label{234-21}
L(l,{\cal C}) = \frac{1}{4\pi} \int  d\tau ds\, \epsilon_{\mu  \nu \rho}  
\frac{dy_\mu}{d\tau} \frac{dx_\nu}{ds} \, \frac{ \bar{x}_\rho(\tau) - x_\rho(s) }{|\bar{x}(\tau)-x(s)|^3}  \;,
\end{equation}
where $x_\mu(s)$ and $\bar{x}_\mu(\tau)$ parametrize $l$ and ${\cal C}$, respectively.

The center elements of $SU(N)$ are given by the $N$th roots of unity
\begin{equation}
\label{245-2D}
z (k) = e^{i k \frac{2 \pi}{N}} \Id \makebox[.5in]{,} k = 0, 1, 2, \ldots, N - 1 \, , 
\end{equation}
where $\Id$ denotes the $N$-dimensional unit matrix. Like all group elements, the center elements can be represented as exponentials of Lie algebra-valued vectors
\begin{equation}
\label{250-2D}
z (k) = e^{i 2 \pi \nu (k)} \, ,
\end{equation}
where $\nu (k)$ are the co-weights, which live in the Cartan subalgebra (with generators $H^a$ in the fundamental representation)
\begin{equation}
\label{255-d}
\nu (k) = \nu^a (k) H_a 
\end{equation}
and define the corners of the fundamental domain (Weyl alcove) of the $SU(N)$ algebra. From eq. (\ref{cele}) it is clear that a center vortex is connected with a non-trivial center element $z (k \neq 0)$. Since $z (k) = z (1)^k$, there exist vortex branching for $SU (N \geq 3)$, which we will, however, not consider in the present paper. Therefore, in the following we will consider only vortices connected with $z (1)$. This is sufficient for SU(2) gauge theory. 

Our objective is to compute the large distance behavior of the ensemble average $ \langle \mathfrak{z} ({\cal C}) \rangle$, summing over all possible numbers and shapes of {\it closed} vortex worldlines, after equipping them with appropriate physical properties.

Initially, we can rewrite (\ref{234-21})
\begin{equation}
\label{265-2}
2 \pi \nu L(l,{\cal C}) = \oint\limits_C d x_\mu A^l_\mu (x) 
\end{equation} 
in terms of a gauge field 
\begin{equation}
\label{270-5}
A_\mu^l (x) =  \frac{\nu}{2}  \oint_l   dx_\nu \,  \epsilon_{\mu  \nu \rho} \,\frac{ x_\rho -x_\rho(s) }{|
	x-x(s)|^3}  \;.
\end{equation}
Then the contribution of the center vortex loop $l$ to the Wilson loop $C$ in eq. (\ref{cele}) becomes
\begin{equation}
\label{275-GX1}
\mathfrak{z} (C) = e^{i \oint\limits_C d x_\mu A^l_\mu} \, .
\end{equation}
Using Stoke's theorem in eq. (\ref{265-2}) 
\begin{equation}
\label{276-2}
\oint_C d x_\mu  A^l_\mu (x)  = \il_{S (C)} d s_\mu B^l_\mu (x) \, ,
\end{equation}
where
\begin{equation}
\label{280-2}
B^l_\mu (x) = \epsilon_{\mu  \nu \rho} \partial_\nu A^l_\rho \;,
\end{equation}
the linking number can be expressed as the intersection number between $l$ and a surface $S({\cal C})$ bordered by ${\cal C}$.
Note that $A^l_\mu (x)$ and $B^l_\mu (x)$ represent the gauge potential and the dual field strength of the center vortex localized on $l$. Indeed from eq. (\ref{270-5}) follows 
\begin{eqnarray}
\label{292-10}
B^l_\mu(x) = 2 \pi \nu  \int_0^L  ds\, u_\mu(s)\, \delta^{(3)}(x-x(s)) \;,
\end{eqnarray}
where we adopted $s$ as the arc length parameter of the center vortex and defined
\begin{equation} 
u_\mu=\frac{dx_\mu}{ds}  \makebox[.5in]{,} u_\mu\in S^2 \;.
\label{ace}
\end{equation} 
Defining a vector field $J^C_\mu(x)$ localized on $S({\cal C})$
\begin{equation}
\label{297-12}
J^C_\mu(x) =  \frac{\nu}{2} \int_{S({\cal C})} ds_\mu \, \delta^{(3)}(\bar{x}(\sigma_1, \sigma_2)-x) \makebox[.5in]{,} 
ds_\mu = d\sigma_1 d\sigma_2\,  \epsilon_{\mu \nu \rho} \, \frac{\partial \bar{x}_\nu}{\partial \sigma_1} 
\frac{\partial \bar{x}_\rho}{\partial \sigma_2} \;,
\end{equation} 
where $\bar{x}(\sigma_1 , \sigma_2)$ is a parametrization of $S({\cal C})$, 
the contribution
of a center vortex $l$ belonging to the center element $z = e^{i 2 \pi \nu} = e^{i 2 \pi k/N}$ can be written as
\begin{equation}
\label{305-5}
e^{i \il^L_0 d s\, u_\mu (s) J^C_\mu (x (s))} =  e^{i  \oint\limits_l   d x_\mu  
J^C_\mu}\, . 
\end{equation}
Since the linking number $L (l, C)$ is symmetric with respect to the interchange of the loops, $L (l, C) = L (C, l)$,  performing this change in Eq. \eqref{305-5} and comparing with Eq. (\ref{275-GX1}),  it follows that the vector fields $A^l_\mu (x)$ and $J^l_\mu (x)$   have to be equivalent.  
Indeed, $A^l_\mu (x)$ (\ref{270-5})  can be gauge transformed to $J^l_\mu$ (\ref{297-12}) \cite{Engelhardt:1999xw,Reinhardt:2001kf}. 
(The former satisfies $\partial_\mu A^l_\mu (x) = 0$). 
Hence the vector field $J^C_\mu (x)$ (\ref{297-12}) represents the gauge potential of a center vortex whose worldline is given by $C$ (instead of $l$).

For a set of $n$ center vortex wordlines 
$l_k$, $k=1, \dots, n$, , parametrized  by $x^{(k)}(s_k)$, $s_k \in [0, L_k]$, the contribution to the Wilson loop is,
\begin{equation}
e^{i\, \sum_{k=1}^n \int_0^{L_k} ds_k\, u^{(k)}_\mu J^C_\mu(x^{(k)}) } \;.
\label{exp}
\end{equation}
In order to identify observables in the  center vortex ensemble with those in the  effective field description to be derived below, it will be convenient to proceed with a general $J_\mu$.  Furthermore, since the coweights $\nu = \nu^a (k) H_a$ occur in the following equations only in the form $e^{i 2 \pi \nu} = e^{i 2 \pi k /N}$ we can  replace below $\nu$ by $k/N$, so that the field $J_\mu (x)$ will no longer be algebra-valued. In addition, we will confine ourselves to a single vortex type $k = 1$, which is sufficient for the gauge group SU(2).  

The center vortex ensemble obtained in $D = 4$   Euclidean Yang-Mills theory after center projection can be modelled by vortices that are distributed according to an action which contains the vortex length and its curvature    
\cite{EngRei00a,Engelhardt:2003wm,Quandt:2004gy}. We therefore describe the intrinsic properties of the center vortices (i.e. $n$ center vortex loops $x^{(k)} (s_k), k = 1, 2, \ldots, n$) by the action
\begin{equation}
S_n^0=\sum_{k=1}^{n}{\int_{0}^{L_{k}} ds_k\, \biggl[ \mu + \frac{1}{2\kappa}\, \frac{du^{(k)}_\mu}{ds}\frac{du^{(k)}_\mu}{ds}  \biggr]}\;.
\label{Szero}
\end{equation}
Here $\mu$ is the tension of the center vortices, i.e. the action cost per length,  and $1/\kappa$ measures their stiffness. The larger $\kappa$ the more flexible are the vortex loops since a smaller penalty is given for the curvature $\dot{u}_\mu \dot{u}_\mu$.
 As is well-known, a finite stiffness $1/\kappa$ is crucial to get a well-defined continuum limit when the worldlines are thought of as polymers, and are discretized in terms of monomers. Regarding the tension parameter,
positive and negative $\mu$ favors small and very long (percolating) vortices in the ensemble, respectively. 
In ref. \cite{Engelhardt:1998wu} it was also shown that the center vortices in YMT do interact and their interaction scales properly in the continuum limit. Therefore we give the center vortex loops a binary interaction of the form
\begin{equation}
S_n^{\rm int}=\frac{1}{2} \sum_{k,k'} \int_{0}^{L_k} \int_{0}^{L_k'} ds_k\, ds_{k'}\,
V\left(x^{(k)}(s_k),x^{(k')}(s_{k'})\right) \;, 
\label{Sint}
\end{equation}
resulting in the partition function
\begin{eqnarray}
Z[J_\mu ]  =  \sum_n  \int [Dl]_n \,  e^{i\, \sum_{k=1}^n \int_0^{L_k} ds_k\, u^{(k)}_\mu J_\mu(x^{(k)}) } 
e^{-\left[ S_n^0 +S_n^{\rm int}\right]} \; .
\label{avez}   
\end{eqnarray}
Here the measure $[Dl]_n$ integrates over all the possible realizations of $n$ center vortices and will be specified later, see eq. (\ref{414-26}) below. 
The average of the Wilson loop for the center vortex ensemble is obtained from
\begin{equation}
\langle \mathfrak{z}({\cal C}) \rangle = \frac{Z[J^C_\mu]}{Z[0]}  \;,
\label{cen-ave}
\end{equation} 
with $J^C_\mu$ given by eq. (\ref{297-12}).
We shall consider repulsive contact interactions which account for excluded volume effects,  $ V(x-y)= (1/\zeta)\, \delta(x-y)$, $\zeta > 0$. 
Then the interaction term of the action can be expressed by the scalar vortex density
\begin{equation}
\rho(x) = \sum_{k}{\int_{0}^{L_{k}} ds_k\, \delta\left(x- x^{(k)}(s_k)\right)} 
\end{equation}
yielding
\begin{equation}
\label{361-6G}
S_{int} = \frac{1}{2 \zeta} \int d^3 x \rho^2 (x) \, .
\end{equation}
It is convenient to linearize this term by means of a scalar field $\phi (x)$ 
\begin{equation}
e^{-\frac{1}{2\zeta} \int d^3x\, \rho^2(x) }=\int [D\phi]\, e^{-W[\phi]}\, e^{i\int d^3x\, \rho(x)\phi(x)}\;,
\label{represent}
\end{equation}
\begin{equation}
W[\phi] =\frac{\zeta}{2}\int d^{3}x\, \phi^2(x) \;.
\label{rhod}
\end{equation} 
Then, we arrive at
\begin{eqnarray}
\lefteqn{ Z[J_\mu] =   \int [D\phi]\, e^{-W[\phi]}  }\nonumber \\
&& \times \, \sum_n\,  \int [Dl]_n\,  e^{ \left[ {    - \sum_{k=1}^{n}\int_0^{L_{k}} ds_k \,
		\left( \omega(x^{(k)}, u^{(k)}) +\frac{1}{2\kappa}\, \dot{u}^{(k)}_\mu \dot{u}^{(k)}_\mu\right) }\right] } \;, 
\label{zeta}  
\end{eqnarray}
where
\begin{equation}
\omega (x,u)  =  \mu - i\phi(x)   - i u_\mu\, J_{\mu}(x)  \;,  
\label{ome}
\end{equation} 
and the measure $[Dl]_n$ is given by,
\begin{eqnarray}
\lefteqn{ [Dl]_n \equiv \frac{1}{n!}\int_{0}^{\infty}\;
	\frac{dL_{1}}{L_{1}}\frac{dL_{2}}{L_{2}}\dots\frac{dL_{n}}{L_{n}} \;
	\int\; dv_1 dv_2 \dots dv_n }  \nonumber \\
&&  \int  [Dv(s_1)]_{v_1,v_1}^{L_1}   \dots [Dv(s_n)]_{v_n,v_n}^{L_n} \makebox[.5in]{,}  dv = d^{3}x  d^2u \;.
\label{414-26}
\end{eqnarray} 
Here the variables $x$, $u$ were collectively denoted by $v$, and $ [Dv(s)]_{v,v}^{L}$ integrates over center vortex worldlines of fixed length $L$ starting and ending at the same position $x$ with the same initial and final tangent vector $u$, which coresponds to smooth closed loops. In eq. (\ref{zeta}) the summation over the number of vortices can be carried out explicitly yielding
\begin{eqnarray}
Z[J_\mu] =   \int [D\phi]\, e^{-W[\phi]} \, e^{  \int_{0}^{\infty}\frac{dL}{L}\;  \int  dv \,  q(v,v,L)  }  \;,   
\label{q-smooth} 
\end{eqnarray} 
where
\begin{equation}
q(v,v_0,L) =  \int  [Dv(s)]_{v,v_0}^L \, e^{-  \int_0^{L} ds \,   
	\left[ \omega(x(s),u(s)) + \frac{1}{2\kappa}\, \dot{u}_\mu \dot{u}_\mu \right] } \;,
\label{discv}
\end{equation}
is the end-to-end probability for a worldline of length $L$  to start at $x_0$, with tangent $u_0$, and end at $x$ with $u$.  $\int d u q (v, v, L)$ represents the partition function of a single closed vortex line of fixed length $L$. Although we will consider a general $N$, for $N\geq 3$ center vortex ensembles also involve fusion rules, conserving the topological charge. This type of branching was not included in the discussion above. That is, in the Yang-Mills context, our vortex model will be particularly appropriate to describe vortex ensembles in $SU(2)$ Yang-Mills theory. Further comments about this point will be given in section 
\ref{discussion}.   

\section{Ensemble average}\label{sectIII}

To perform the ensemble average, we shall closely follow the calculations given in Refs. \cite{deLemos:2011ww} and \cite{Oxman:2014hla}, based on polymer techniques developed in Refs. \cite{PhysRevLett.73.3235} and \cite{fredrickson2005equilibrium}, which are briefly reviewed here. 

In the functional integral \eqref{discv}, all the paths have fixed length $L$ so they can be discretized in terms of $M$ small segments (``monomers'') of length $\Delta L = L/M$ between the points $x_j$ and $x_{j+1}$, $j=0, \dots, M-1$.  Naming $x=x_{M}$, $u=u_{M}$ and defining,
\begin{eqnarray}
\lefteqn{q_M(x,u,x_0,u_0)=\int d^3x_{1} d^2u_{1}\dots d^3x_{M-1} d^2u_{M-1}\,}  \nonumber \\
&&\times  \, e^{-\Delta L \,\sum_{i=1}^{M}\omega(x_{i}, u_{i})} \prod_{j=0}^{M-1} \psi(u_{j+1}-u_{j})
\delta(x_{j+1} - x_{j} - u_{j+1}\Delta L)\;,
\label{9}
\end{eqnarray}
\begin{equation}
\psi(u-u')=\mathcal{N}\, e^{-\frac{1}{2\kappa}\Delta L\left(\frac{u-u'}{\Delta
		L}\right)^{2}} \makebox[.5in]{,} \int d^2u' \, \psi(u-u')=1\;,
\end{equation}
we have,
\begin{equation}
q(v, v_0,L) = \lim_{M\to \infty} q_M(x,u,x_0,u_0)  \;.
\end{equation} 
Now, separating the integral over $d^3x_{M-1} d^2u_{M-1}$ in Eq. (\ref{9}) and renaming  $x'=x_{M-1}$, $u'=u_{M-1}$, one finds
\begin{eqnarray}
\lefteqn{q_{M}(x,u,x_0,u_0)=\int  d^3x' d^2u'\, }  \nonumber \\
&& \times \, e^{-  \omega(x, u) \Delta L} \psi(u-u') \,
\delta(x - x' - u\Delta L) \, q_{M-1}(x',u',x_0,u_0) \;.
\end{eqnarray}
where $q_{M-1}(x',u',x_0,u_0)$ is the end-to-end probability for a line with initial condition $x_0$, $u_0$ and lenght $L - \Delta L$, formed by  
$M-1$ monomers, to end at $x'$, $u'$. Integrating over $x'$ and using the notation of the continuum, with an infinitesmial $\Delta L$, we find
\begin{eqnarray}
\lefteqn{q(x,u,x_0,u_0,L)= }  \nonumber \\
&& \int  d^2u'\, e^{-  \omega(x, u) \Delta L} \psi(u-u' ) \, q (x-u \Delta L,u',x_0,u_0,L-\Delta L) \;.
\label{recu}
\end{eqnarray} 
For a finite $\kappa $,  the terms first order in $\Delta L$ lead to, 
\begin{equation}
\partial_{L}q = \left[-\mu + \frac{\kappa}{2}\, \hat{L}^{2}_{u} + i\phi(x) -u_{\mu} D_\mu \right]q
\makebox[.5in]{,} D_\mu = \partial_{\mu}- i J_\mu   \;,
\label{difu}
\end{equation}
with the initial condition,
\begin{eqnarray}
q(x,u, x_0,u_0,0)= \delta(x-x_{0})\, \delta(u-u_{0}) \;.
\label{iniab1}
\end{eqnarray}
The operator $ \hat{L}^{2}_{u}$ is the Laplacian on the sphere $u_\mu \in S^2$. It arises from expanding $q (\ldots, u', \ldots)$
in the integrand of Eq. \eqref{recu}  in powers of $u'-u$,  and computing the moments of the distribution $\psi(u-u')$.\footnote{The quantity $q (x, u, x_0, u_0; L)$ can be interpreted as an Euclidean transition amplitude for the evolution during the Euclidean "time" interval $L$ and eq. (\ref{difu}) is nothing but the corresponding imaginary time-dependent Sch\"odinger equation. In fact, the derivation of eq. (\ref{difu}) proceeds analogously to the derivation of the time-dependent Schr\"odiger equation from the functional integral in quantum mechanics, see \cite{feynman1965quantum}.} As $u_\mu$ carries angular momentum $l=1$, in an expansion of $q$ in terms of spherical harmonics,
Eq. \eqref{difu} couples the different $l$-sectors 
\begin{equation}
q(x,u, x_0,u_0,L)=\sum_{l=0} {\cal Q}_{l} (x,u,x_0,u_0,L) \;, 
\end{equation}
\begin{equation}
{\cal Q}_{l} (x,u,x_0,u_0,L) = \sum_{m=-l}^{l}  {\cal Q}_{lm}
(x,x_0,u_0,L)\, Y_{lm}(u)  \;,
\label{l2}
\end{equation}
\begin{equation}
\hat{L}^{2}_{u}\, Y_{lm}(u)=  -l(l+1)\, Y_{lm}(u)\;.
\end{equation} 
Using the completeness relation,
\begin{eqnarray}
\label{477-40}
\delta (u-u_0)=  \sum_{l,m}  Y^\ast_{lm}(u_0)\,  Y_{lm}(u)\;,
\end{eqnarray}
the initial condition now reads,
\begin{equation}
{\cal Q}_{lm} (x,x_0,u_0,0) = \delta(x-x_{0})\,
Y^\ast_{lm}(u_0) \;.
\label{ini-lm}
\end{equation}
For small stiffness and large $L$, the correlations between the initial tangent direction $u_0$ and the final one $u$ become small, thus favouring small angular momenta. In the semiflexible limit, which corresponds to a finite but small stiffness, the angular momenta $l\geq 2$ can be disregarded in the expansion (\ref{difu}). In this case, the solution to eq. (\ref{difu}) can be approximated by \cite{PhysRevLett.73.3235},   
\begin{equation}
q(x,u,x_0,u_0,L) \approx   {\cal Q}_{0} (x,u,x_0,u_0,L) +  {\cal Q}_{1} (x,u,x_0,u_0,L) \;,
\label{l1}
\end{equation}
\begin{equation}
O {\cal Q}_{0}  + \partial_{L}{\cal Q}_{0} \approx 0 \makebox[.5in]{,} O = - \frac{1}{3 \kappa}\, D_\mu D_\mu  +\mu -i \phi 
\;,
\label{eq0}
\end{equation}
\begin{equation}
{\cal Q}_{1}  \approx -\frac{1}{\kappa} \, (u\cdot D)\,{\cal Q}_{0} \;.
\label{l3}
\end{equation}
Indeed, using Eqs. \eqref{l1}-\eqref{l3}, it can be easily verified that
\begin{eqnarray}
\lefteqn{\partial_L q +  \left[\mu - \frac{\kappa}{2}\, \hat{L}^{2}_{u}   - i\phi(x) + ( u \cdot D) \right]q }
\nonumber \\ 
&  & \approx    -\frac{1}{\kappa} \, \left[ u_{\mu}\, u_{\nu}- (1/3)\, \delta_{\mu\nu} \right] D_\mu D_\nu  {\cal Q}_{0} \;,
\end{eqnarray}
where the second member involves an operator carrying angular momentum $l=2$ (a traceless symmetric tensor), which was disregarded in the ansatz \eqref{l1}. Then, the equations close when restricted to the $l=0,1$ sectors.
For a discussion involving the coupled equations for the whole tower of angular momenta, see Refs. \cite{deLemos:2011ww} and \cite{Oxman:2014hla}. Summarizing, keeping the dominant term $ {\cal Q}_{0}$, and using Eq. 
\eqref{eq0}, with the $l=0$ initial condition in Eq. \eqref{ini-lm}, 
\begin{equation}
{\cal Q}_{0} (x,u,x_0,u_0,0) = \delta(x-x_{0})\,
Y^\ast_{00}(u_0)\, Y_{00}(u) =
\frac{1}{4\pi}\, \delta(x-x_{0})\;,
\label{l30}
\end{equation} 
the end-to-end probability turns out to be,
\begin{equation}
q(x,u,x_0,u_0,L) \approx  \frac{1}{4\pi}  \langle x| e^{-LO} | x_0 \rangle\;.
\label{e2e}
\end{equation} 
Inserting this expression into Eq. (\ref{q-smooth}), and using  
\begin{equation}
\label{514-F10}
\il^\infty_0 \frac{d L}{L} \int d^3 x \langle x | e^{- L O} | x \rangle = \int \frac{d L}{L}\,  Tr\,  e^{- L O} = Tr \ln O  \;,
\end{equation}
\begin{equation}
\label{512-F11}
e^{- Tr \ln O} = (Det \, O)^{- 1} = \int \cD V \cD \bar{V} e^{- \int d^3 x\, \bar{V} O V} \;,
\end{equation}
we find the following representation of the partition function for center vortices with small but non-zero stiffness $1/\kappa$ 
\begin{eqnarray}
Z[J_\mu ]&\approx &\int [D\phi]\, e^{-W[\phi]} \int [DV[D\bar{V}]\; e^{-\int
	d^{3}x\, \bar{V} O V }\nonumber 
\nonumber \\
&= & \int [DV[D\bar{V}] \, e^{-\int
	d^{3}x\,\left[ \frac{1}{3 \kappa}\,  \overline{D_\mu V} D_\mu V + \mu\, \overline{V} V + \frac{1}{2\zeta}\, (\overline{V} V)^2 \right]}  \;.
\label{zetaeff}
\end{eqnarray}
To exhibit the physical meaning of the complex field $V$ in terms of the initial center vortex ensemble, we express $Z[J_\mu]$ in Eq. \eqref{avez} using the total dual field strength $B_\mu$ of the thin 
center vortices, (c.f. eq. (\ref{292-10}))  
\begin{eqnarray}
B_\mu (x) = \sli^n_{k = 1} B^{l_k}_\mu (x) = \frac{2 \pi}{N} \sum_{k=1}^n  \int_0^{L_k}  ds_k \, u^{(k)}_\mu(s_k)\, \delta^{(3)}(x-x^{(k)}(s_k)) \;,
\label{efemu}
\end{eqnarray}
and include an external source $J(x)$ to the scalar vortex density $\rho (x)$,
\begin{eqnarray}
Z [ J, J_\mu ]  
=\sum_n  \int [Dl]_n \,  
e^{-\left[ S_n^0 +S_n^{\rm int}\right]}  \, e^{ i \int d^3x\, \rho(x)  J(x)}   \, e^{ i \int d^3x\, B_\mu(x) J_\mu(x)} 
\;.
\label{zjj} 
\end{eqnarray}  
After linearizing the interaction term, this amounts to the substitution $\phi(x) \to \phi(x) + J(x)$ in 
Eq. \eqref{ome}. Then, following the same steps that led to Eq. (\ref{zetaeff}), with $O \to O - iJ(x)$ (cf. 
Eq. \eqref{eq0}), we obtain
\begin{eqnarray}
\lefteqn{ Z[J, J_\mu ]
	=  \int [D\phi]\, e^{-W[\phi]} \int [DV[D\bar{V}]\; e^{-\int
		d^{3}x\, \bar{V} (O -iJ) V } }
\nonumber \\
&& =  \int [DV[D\bar{V}] \, e^{-\int
	d^{3}x\,\left[ \frac{1}{3 \kappa}\,  \overline{D_\mu V} D_\mu V + \mu\, \overline{V} V + \frac{1}{2\zeta}\, (\overline{V} V)^2 \right]}  \, e^{ i \int d^3x\, \bar{V}V(x)  J(x)}     \;.
\label{zetajd}
\end{eqnarray}
Taking functional derivatives in Eqs.  
(\ref{zjj}) and (\ref{zetajd}), with respect to $J(x_1), J(x_2), \dots$ and $J_{\mu_1}(x_1), J_{\mu_2}(x_2), \dots$ at non-coinciding points, and setting the external sources to zero (including the sources $J_\mu (x)$ induced by the Wilson loop), we find the following correspondence between correlation functions (in the absence of the loop ${\cal C}$),
\begin{equation}
\langle \rho(x_1) \rho(x_2) \dots \rangle \longleftrightarrow \langle \bar{V} V (x_1) \, \bar{V}V(x_2) \dots \rangle \;,
\label{rhoden}
\end{equation}
\begin{equation}
\langle B_{\mu_1}(x_1) B_{\mu_2}(x_2) \dots \rangle
\longleftrightarrow \langle K_{\mu_1} (x_1) K_{\mu_2}(x_2) \dots \rangle \;,
\nonumber
\end{equation}
where
\begin{equation}
K_\mu=  \frac{2 \pi}{6 N \kappa} ( V \partial_\mu \bar{V} - \bar{V} \partial_\mu V )  \;.
\label{Kcorr} 
\end{equation} 
Since reversing the orientation of the vortex flux changes the sign of $B_\mu (x)$, correlation functions with an odd number of $B_\mu (x)$'s vanish. 

\section{XY and Villain models with $J_\mu (x)$}\label{sectIV} 

In the previous sections, we have obtained the effective field representation for the average of center elements $\langle \mathfrak{z}({\cal C}) \rangle $ given by Eqs. \eqref{cen-ave} and \eqref{zetaeff}. On the other hand, in the
initial center vortex ensemble this average is represented by,
\begin{eqnarray}
&& \langle \mathfrak{z}({\cal C}) \rangle = \frac{1}{\cal N} \sum_n  \int [Dl]_n \,  \cos \left(  \il_C dx_\mu B_\mu (x) \right) e^{-\left[ S_n^0 +S_n^{\rm int}\right]} \;,
\end{eqnarray}
\begin{equation}
\mathcal{N} = \sum_n  \int [Dl]_n \,  e^{-\left[ S_n^0 +S_n^{\rm int}\right]} 
\end{equation}
(cf. Eqs. (\ref{avez}) and (\ref{efemu})). Again, we have used the fact that vortex configurations come in pairs of opposite orientation, $B_\mu$ and $-B_ \mu$, to get an explicitly real expression, in accordance with the real integrand in Eq.  
(\ref{zetaeff}). This field representation is valid for semiflexible vortices (small but nonzero values of $1/\kappa$), a condition that has permitted us to keep only the smallest angular momenta in the tangent $u$-space and to obtain a quadratic kinetic term. 
  
If the initial Yang-Mills theory were coupled to a set of Higgs fields, such that the center vortices emerged as classical saddle points of the action, the parameter $\mu$ would be positive. This action cost would lead to an ensemble of small loops, and a  
perimeter law for large Wilson loops, see sect. \ref{sectV}.  Let us analyze this situation from the point of view of the effective field theory. When $\mu > 0$, the functional integral over the complex field $V$ in Eq. \eqref{zetaeff} can be
computed perturbatively taking as reference the quadratic Lagrangian
\begin{equation}
{\cal L}_0 = \frac{1}{3 \kappa}\,  \overline{D_\mu V} D_\mu V + \mu\, \overline{V} V  \;.
\end{equation}
Then the partition function is dominated by the functional determinant of the inverse propagator of the massive field $V$,
with squared mass $3 \kappa \mu > 0$,
\begin{equation}
Z_0 [J_\mu]\sim \exp{ \left[ - \ln {\rm Det} (-D_\mu D_\mu + 3 \kappa \mu) \right] } \;. 
\end{equation}
Recalling that the complex field $V$ is minimally coupled to the ``gauge field'' $J_\mu$ 
(cf. Eq. \eqref{difu}), and that the effective action is gauge invariant, the partition function $Z_0 [J_\mu]$ can only depend on the ``field strength'' $\epsilon_{\mu \nu \rho} \partial_\nu J_\rho$. That is, the average $\langle \mathfrak{z}({\cal C}) \rangle $, where $J_\mu = J^C_\mu$, depends on $\epsilon_{\mu  \nu \rho} \partial_{\nu} J^C_\rho$.
Now, from the definition of $J^C_\mu (x)$ (\ref{297-12}), it follows that
\begin{equation}
\label{606-10}
\epsilon_{\mu  \nu \rho} \partial_{\nu} J^C_\rho (x) = \frac{2\pi}{N} \oint\limits_C d \bar{x}_\mu \, \delta^{(3)} (x - \bar{x}) \;,
\end{equation}
which is localized on the Wilson loop ${\cal C}$. This, together with the nontrivial mass scale $3 \kappa \mu > 0$ implies  
a perimeter law in the Higgs phase. 

On the other hand, from lattice simulations, we know that center vortices percolate in the pure Yang-Mills vacuum and that, from the ensemble point of view, this leads to an area law for Wilson loops. Percolated vortices are necessary large and  require $\mu < 0$ in the vortex action. Let us now investigate how in this case the area law emerges in the effective field description. For $\mu < 0$ we have for the potential term in the effective field theory (\ref{zetaeff})
\begin{eqnarray}
\label{614-51}
\mu \bar{V}V +  \frac{1}{2\zeta}\, (\overline{V} V)^2  = 
\frac{1}{2\zeta}\, (\overline{V} V -v^2)^2 -v^4/2\zeta  \;,
\end{eqnarray} 
with $v^2 = -\mu \zeta > 0$. This potential breaks the underlying global U(1) symmetry spontaneously. Dropping the irrelevant constant in eq. (\ref{614-51}), Eq. \eqref{zetaeff} becomes,
\begin{eqnarray}
Z [J_\mu]
\approx \int [DV[D\bar{V}] \, e^{-\int
	d^{3}x\,\left[ \frac{1}{3 \kappa}\,  \overline{D_\mu V} D_\mu V + \frac{1}{2\zeta}\, (\overline{V} V -v^2)^2 \right]}  \;.
\label{effmod} 
\end{eqnarray} 
From the equivalence $\langle \rho (x) \rangle \leftrightarrow \langle \bar{V} V (x) \rangle$ established in eq. (\ref{rhoden})  we find that in the vacuum of the effective theory (\ref{477-40}) the center vortices are condensed having a scalar density $\langle \rho (x) \rangle \sim v^2$.
For an evaluation of $\langle \mathfrak{z} (C) \rangle$ we have to include quantum fluctuations around the classical vacuum configuration $\bar{V} V = v^2$. For this purpose we write the unitary field $V (x)$ as 
\begin{equation}
V(x) = \rho(x)\, e^{i\gamma(x)} \makebox[.5in]{,} \rho(x)= v + h(x) \;.
\end{equation}   
For sufficiently weak vortex interactions the potential in eq. (\ref{effmod}) tolerates only small fluctuations in the field $h (x)$, while $\gamma (x)$ is a Goldstone field, whose fluctuations are not restricted by the potential. Furthermore, $\gamma (x)$  is a compact field defined modulo $2 \pi$. 
To have a well-defined description of the soft degrees of freedom $v\, e^{i\gamma}$, and to keep their compact character, we switch to the lattice version of  Eq. (\ref{effmod}),
\begin{equation}
S_{\rm latt} = \sqrt{\eta} \, \left|\partial_\mu e^{i\gamma} - iJ_\mu \,  e^{i\gamma}\right|^2_{\rm latt}  \makebox[.5in]{,} \sqrt{\eta} = \frac{v^2}{3\kappa} \; ,  
\end{equation}  
where we have ignored the small amplitude fluctuation field $h (x)$, putting $\rho (x)$ to its vacuum value $v$.  
Note that  the relevant dimensionful parameter here is not $v^2$ (scalar density) but $\sqrt{\eta} = v^2/3\kappa $, which  also controls  the {\it vector} vortex (current) density (cf. Eq. (\ref{Kcorr})).  Then, this parameter is expected to be related to the number of vortices intersecting a given surface per unit area, that for dimensional reasons should scale as $\sim \eta^2$. 

For a cubic lattice with $M$ sites  $\mathbf{x}$, spacing $a$, and oriented links $\hat{\mu}$, the discretized covariant derivative is,
\begin{equation}
(\partial_\mu -i J^C_\mu)V \longrightarrow \frac{1}{a}\, (e^{i\gamma(\mathbf{x} + \hat{\mu})}- U_\mu(\mathbf{x})\, e^{i\gamma(\mathbf{x}) } )   \makebox[.5in]{,} U_\mu(\mathbf{x}) = e^{i \alpha_\mu (\mathbf{x})}\;.
\end{equation}  
When $J^C_\mu$ is smooth, $\alpha_\mu (\mathbf{x}) =  a J^C_\mu (\mathbf{x}) $. From the explicit form of $J^C_\mu$ (\ref{297-12}) follows that $ \alpha_\mu (\mathbf{x}) = \frac{2\pi}{N}$ if the surface $S({\cal C})$ is crossed by the link (in the direction of the normal to $S({\cal C})$), and zero otherwise.  Therefore, we are led to the $3${\rm d} XY model with frustration $\alpha_\mu(\mathbf{x})$,
\begin{equation}
S_{\rm latt} = 2\sqrt{\eta}\, a^3 \sum_{\mathbf{x}, \mu } \frac{1}{a^2} [1-  \cos (  \nabla_\mu \gamma(\mathbf{x})   - \alpha_\mu ({\mathbf{x}}))  ] 
\makebox[.3in]{,}   \nabla_\mu \gamma(\mathbf{x})  =   \gamma(\mathbf{x} + \hat{\mu})- \gamma(\mathbf{x})  \;.
\end{equation}
and partition function 
\begin{eqnarray}
&& Z_{\rm XY}(\alpha_\mu) = \prod_{\mathbf{x}}\int_{-\pi}^{+\pi} \frac{d\gamma(\mathbf{x})}{2\pi} \,  e^{\beta \sum_{\mathbf{x}, \mu  }   \cos (  \nabla_\mu \gamma(\mathbf{x})   - \alpha_\mu ({\mathbf{x}}))   }  \;,
\label{zlattn}
\end{eqnarray}
where
\begin{equation}
\beta = 2a \sqrt{\eta}  \;.
\label{be}
\end{equation}
The normalized average of the Wilson loop (\ref{cen-ave})  is given by
\begin{eqnarray}
 \langle \mathfrak{z}({\cal C})  \rangle_{\rm latt} 
	\approx \frac{~Z_{\rm XY}(\alpha_\mu)}{Z_{\rm XY}(0)}   \;,    
	\label{quo}
\end{eqnarray}
In fact, to make contact with the continuum, we are interested in the critical region where the correlation lengths become large with respect to the lattice spacing. 
In the literature, studies about the frustrated $3${\rm d} XY model can be found for specific realizations of
$\alpha_\mu(x)$. The fully frustrated case, with homogeneous frustration vector, has been extensively analyzed (see Ref. \cite{PhysRevB.73.224504} and references therein).  
This vector has $x,y$ and $z$ components given by the circulation of $\alpha_\mu(x)$ along plaquettes on the $yz$, $zx$ and $xy$-planes, respectively. Different frustration vectors have been studied, each one displaying its own critical properties.  
The case of a random phase shift has been discussed in Ref. \cite{PhysRevB.54.16024}. To the best of our knowledge, there are no studies for the $3${\rm d}  XY model with phase shifts localized on a geometric region, as needed to compute the Wilson loop. However, in this case, the phase shift vanishes along the whole lattice but on those links that cross $S({\cal C})$. Moreover, the frustration vector is only nonzero on plaquettes
that contain just one link with nontrivial shift, which are placed at the border of $S({\cal C})$. Then, to analyze $\langle \mathfrak{z}({\cal C})  \rangle_{\rm latt} $ in Eq. \eqref{zlattn},  we shall assume that the thermodynamic properties of the system\footnote{We are using the analogy with classical statistical mechanics where quantum fluctuations in $(2+1)${\rm d} are thought of as ``thermal'' fluctuations in $3${\rm d}.} are those of the problem without frustration, 
that has a critical point at $\beta_c \approx 0.454$ (see  Ref. \cite{kleinert1989gauge} and references therein).

In the critical regime, different models within the same universality class can be used.   Outside this region, at very small (large) $\beta$, which means large (small) quantum fluctuations,  the details are in general model dependent. 
Let as describe what happens when we go from very small $\beta$  to $\beta_c$. In this region, the XY model is in excellent agreement with the Villain model, which is in the same universality class. Let us summarize the main steps underlying this approximation following Ref. \cite{kleinert1989gauge}, where the XY and Villain approximations (without frustration) were extensively reviewed and studied for a superfluid. An expansion in powers of $\beta$ leads to integrals of products of cosine functions. To organize the calculation, it is more convenient to use the Fourier decomposition, 
\begin{equation}
e^{\beta \cos \gamma}  = \sum_{b\, = -\infty}^{+\infty} I_b(\beta) e^{ib\gamma}  \;,
\end{equation}
where $I_b(\beta)$ is the modified Bessel function of integer order $b$. 
Then, introducing integer valued variables $b_\mu(\mathbf{x})$,  integrating by parts on the lattice, and over the $\gamma$-variables yields 
\begin{equation}  
Z_{\rm XY}(\alpha_\mu)  =  (I_0(\beta))^{3M}   \sum_{\{ b_\mu (\mathbf{x}) \} }     \prod_{\mathbf{x}, \mu}   \frac{ I_{b_\mu(\mathbf{x}) }(\beta) }{I_0(\beta)} \,  e^{-i \, b_\mu (\mathbf{x})  \, \alpha_\mu (\mathbf{x})}   \label{Zprime}  \;.
\end{equation} 
The summation over $\{ b_\mu (\mathbf{x})\}$ runs over non-backtracking oriented closed loops of unit strength. On a given link $(\mathbf{x}, \hat{\mu})$, if just one loop passes with the same (opposite) orientation as $\hat{\mu}$, then $b_\mu({\mathbf{x}})$ takes the value $+1$ ($-1$).  If 
$n \geq 2$ loops pass on this link, all with the same orientation, then the variable takes the value $b_\mu({\mathbf{x}}) =\pm n$ depending  on how the loops are oriented with respect to $\hat{\mu}$. These loops are analog to the fluxes of $B_\mu$ through the plaquettes.

 The Villain approximation to $Z_{\rm XY}(\alpha_\mu)$ in Eq. \eqref{zlattn} is given by the replacement,
\begin{equation}
e^{\beta \cos \gamma} \longrightarrow R_{\rm V}(\beta) \sum_{n=-\infty}^{+\infty}
e^{-\frac{\beta_{\rm V}}{2} \, (\gamma -2 \pi n)^2}  \;,
\end{equation}
with
\begin{equation}
R_{\rm V}(\beta) = \sqrt{2\pi \beta_{\rm V}} \,  I_0(\beta) \makebox[.5in]{,} -  \frac{1}{2\beta_{\rm V}(\beta)} =   \ln (I_1(\beta)/I_0(\beta))   \;.
\label{Vprop}
\end{equation}
This leads to
\begin{eqnarray}
\lefteqn{ Z_{\rm XY} (\alpha_\mu) \approx Z_{\rm V} (\alpha_\mu) } \nonumber \\
&& = (R_{\rm V}(\beta))^{3M} \prod_{\mathbf{x}}\int_{-\pi}^{+\pi} \frac{d\gamma(\mathbf{x})}{2\pi}  \sum_{\{ n_\mu(\mathbf{x}) \}}
e^{-\frac{\beta_{\rm V}}{2} \, (\nabla_\mu \gamma -\alpha_\mu(\mathbf{x}) -2 \pi n_\mu({\mathbf{x}}))^2}   \;.
\end{eqnarray}
Next,  the Gaussian weights can be linearized with continuous real fields $C_\mu(\mathbf{x})$. Then, the sum over $n_\mu(\mathbf{x})$ can be carried out explicitly using the Poisson formula.  This replaces the $C_\mu(\mathbf{x})$ integrals by a sum over integers $b_\mu(\mathbf{x})$,
\begin{eqnarray}
Z_{\rm V} (\alpha_\mu) & = & \left( \frac{R_{\rm V}}{\sqrt{2\pi \beta_{\rm V}}} \right)^{3M}
\sum_{\{ b_\mu(\mathbf{x}) \}} e ^{-\frac{1}{2\beta_{\rm V}}  \sum_{\mathbf{x}, \mu} 
( b_\mu (\mathbf{x}))^2  }  \,   e^{ -i \sum _{\mathbf{x}, \mu}b_\mu (\mathbf{x})  \, \alpha_\mu (\mathbf{x})} 
\nonumber \\
&& \times \,  \prod_{\mathbf{x}}\int_{-\pi}^{+\pi} \frac{d\gamma(\mathbf{x})}{2\pi}  \,
e^{ i \sum _{\mathbf{x}, \mu}b_\mu (\mathbf{x})  \, \nabla_\mu \gamma  (\mathbf{x})}  \;.
\end{eqnarray}
Finally, using Eq. \eqref{Vprop}, one finds
\begin{eqnarray}
Z_{\rm V}(\alpha_\mu) 
= (I_0(\beta))^{3M}  \sum_{\{ b_\mu (\mathbf{x}) \} } \prod_{\mathbf{x}, \mu}   \left( \frac{I_1(\beta)}{I_0(\beta)}  
\right)^{ (b_\mu (\mathbf{x}))^2  }  e^{ -i b_\mu (\mathbf{x})  \, \alpha_\mu (\mathbf{x})}  \;,
\label{ZprimeV}  
\end{eqnarray}   
which is to be compared with Eq. \eqref{Zprime}. That is, the Villain approximation amounts to the replacement,
\begin{equation}
\frac{ I_{b_\mu(\mathbf{x}) }(\beta) }{I_0(\beta)} \rightarrow  \left( \frac{I_1(\beta)}{I_0(\beta)} 
\right)^{\sum_{\mathbf{x}, \mu} b_\mu (\mathbf{x}) b_\mu (\mathbf{x}) }  \;.
\label{rep}
\end{equation}

\section{Wilson loop $\beta$-behavior}\label{sectV}
\label{beh}

For definiteness, let us consider a planar Wilson loop ${\cal C}$ and a planar surface $S({\cal C})$ whose normal points along the $\hat{1}$-axis. This surface is placed between the sets of sites $\{ \mathbf{z} \}$ and 
$\{ \mathbf{z} + \hat{1}\}$, that is, it is crossed by the links that run from  
$ \mathbf{z} $ to $\mathbf{z} + \hat{1}$.  
Then, we have,
\begin{equation}
e^{ -i \sum _{\mathbf{x}, \mu}b_\mu (\mathbf{x})  \, \alpha_\mu (\mathbf{x})} 
= e^{ -\frac{2\pi i}{N} \sum _{\mathbf{z}} b_1 (\mathbf{z})  } 
\label{factor} \;.
\end{equation}
At very small $\beta$ (high ``temperatures''),  the first contributions to $Z_{\rm XY}(\alpha_\mu)$ and $Z_{\rm V}(\alpha_\mu)$ coincide, and are of order $\beta^4$. They correspond to loops of length 4, running on the sides of the plaquettes. 
If none of the loop sides is a link that crosses $S({\cal C})$, then $b_1(\mathbf{z})=0$. If two sides of the plaquette cross $S({\cal C})$, they have different orientations with respect to $\hat{1}$, so they do not contribute to $\sum _{\mathbf{z}} b_1 (\mathbf{z}) $. Only loops with just one side crossing $S({\cal C})$ give a nontrivial factor (\ref{factor}).  If $M_P$ is the number of links running from $\mathbf{z}$ to $\mathbf{z}+ \hat{1}$ and placed on the perimeter of  
$S({\cal C})$, then,
\begin{eqnarray}
Z_{\rm XY} \approx Z_{\rm V}\approx (I_0(\beta))^{3M}   \left( \frac{I_1(\beta)}{I_0(\beta)} 
\right)^4  \left[ (6M-2M_P) + 2M_P \cos \frac{2\pi}{N} \right]   \;.
\end{eqnarray}
This leads to the average,
\begin{equation}
\langle \mathfrak{z}({\cal C}) \rangle_{\rm latt} =   1- \frac{2}{3} \frac{M_P}{M} \sin^2 \left(   \frac{\pi}{N} \right)  \;,
\end{equation}
and
\begin{equation}
-\ln \langle \mathfrak{z}({\cal C}) \rangle_{\rm latt} \approx   \frac{2}{3a} \frac{P}{M} \sin^2 \left(   \frac{\pi}{N} \right)   \;,
\end{equation}
where $P = M_P \,a$ is the perimeter of ${\cal C}$. As $\beta$ is increased (the ``temperature'' decreased), keeping away from $\beta_c$, the expansion of the partition function will require higher orders in $\beta$. More powers in $\beta$ imply that larger loops and multiple smaller loops are produced.  In the language of superfluids, loops of superflow are generated as the temperature is decreased toward the critical temperature. 
In our context, more and more center vortices are generated as we approach the continuum limit. Anyway, at any finite order in $\beta$ a perimeter law 
(with a renormalized prefactor) is expected.

Let us now analyze the situation close to $\beta_c$. In what follows, we shall denote the total configuration space 
as $  \{ b_\mu (\mathbf{x}) \} = \{ 0\} \cup\{ B_0 \}$, where $\{ 0 \}$ and $\{ B_0 \} $ represent the trivial, $b_\mu(\mathbf{x}) \equiv 0$, and nontrivial configurations, respectively.  
This is the intial set we shall consider to perform a sequence of approximations.  For any subset $\{ A \} \subset  \{ B_0 \}$,  we define 
\begin{eqnarray}
F_{\{ A\}}(\alpha_\mu) = \sum_{\{ A \} } \prod_{\mathbf{x}, \mu}   \left( \frac{I_1(\beta)}{I_0(\beta)}  
\right)^{ (b_\mu (\mathbf{x}))^2  }  e^{ -i b_\mu (\mathbf{x})  \, \alpha_\mu (\mathbf{x})}  \; .
\end{eqnarray} 
We are interested in computing,
\begin{eqnarray}
\label{652-14}
Z_{\rm V}(\alpha_\mu) 
= (I_0(\beta))^{3M} (1+ F_{\{ B_0\}}(\alpha_\mu) ) \makebox[.3in]{,} \langle \mathfrak{z}({\cal C})  \rangle_{\rm latt} 
	\approx \frac{~Z_{\rm V}(\alpha_\mu)}{Z_{\rm V}(0)} = \frac{(1+ F_{\{ B_0 \}}(\alpha_\mu) )}{(1+ F_{\{ B_0\}}(0) )}  \;.
\end{eqnarray} 
The space $\{ B_0 \}$ can be partitioned into three disjoint subsets: $\{ A_{0} \}$, where none of the loops cross $S(\mathcal{C})$ ($b_1(\mathbf{z}) \equiv 0$); $\{ B_1 \}$, where all the loops cross $S(\mathcal{C})$ at least once, and the rest $\{ R_0 \} $, where configurations contain at least one loop of each type. 
That is,
\begin{eqnarray}
F_{\{ B_0 \} } & =  &  F_{\{ A_0 \}} + F_{\{ B_1 \}} + 
F_{\{ R_0 \} } \;.
\label{appF} 
\end{eqnarray}    
Close to the transition point, it is well-known that smaller loops with higher fluxes $\pm2, \pm3, \dots$ are irrelevant with respect to larger loops with unit flux, due to the difference in configurational entropy \cite{kleinert1989gauge}. The replacement in  Eq. \eqref{rep} is exact for configurations $\{ B_0\}' $ characterized by $b_\mu(\mathbf{x}) = \pm 1,$ or $ 0$ ($I_{-1}(\beta)= I_1(\beta)$). This, together with the excellent agreement between the XY and Villain models, indicates that loop configurations that meet at a link
are irrelevant with respect to those with single occupation. This  refers to loops with the same orientation; those that meet with opposite orientations were already forbidden 
for non-backtracking loops. This is the way the initial properties of the ensemble are  encoded in the statistical properties of the Villain model close to $\beta_c$. Negative tension and positive stiffness is now related with the preference for larger non-backtracking loops to the detriment of smaller ones. In addition, the statistical irrelevance of multiple occupation or, in other words, excluded volume effects, can be traced back to the repulsive interactions. Then. the calculation can be approximated by 
\begin{eqnarray}
F_{\{ B_0 \} } \approx F_{\{ B_0  \}' } =   F_{\{ A_0 \}'} + F_{\{ B_1 \}'} + 
F_{\{ R_0 \}' } \;,
\label{appFp}
\end{eqnarray}  
with $ \{ A_0 \}$, $\{ B_1 \} $, $ \{ R_0 \}$    in Eq. \eqref{appF} replaced by 
$ \{ A_0 \}'$, $\{ B_1 \}' $, $ \{ R_0 \}'$  only keeping the relevant loops.
For this type of configuration, we can replace $(b_\mu (\mathbf{x}))^2 \to |b_\mu (\mathbf{x})|$, 
\begin{eqnarray}
F_{\{ A\}'}  = \sum_{\{ A \}' } e ^{-  \sum_{\mathbf{x}, \mu} 
\left( \frac{1}{2\beta_{\rm V}} \, |b_\mu (\mathbf{x})| + i   \, b_\mu (\mathbf{x})  \, \alpha_\mu (\mathbf{x}) \right)  }  \;.
\label{main}
\end{eqnarray}
Up to a factor $a$, $\sum_{\mathbf{x}, \mu} |b_\mu (\mathbf{x})|$ adds the loop lengths, while $\sum_{\mathbf{x}, \mu} b_\mu (\mathbf{x})  \, \alpha_\mu (\mathbf{x}) $ adds their linking numbers. Then, for two subsets  $\{ A\}'$ and $\{ B\}'$ for which the loop combinations $\{ A\}' \times \{ B\}'$ do not share any link, we have 
\begin{eqnarray}
F_{\{ A \}' \times \{ B \}'} = F_{\{ A \}' } \, F_{\{ B \}' }  \;.
\end{eqnarray}
Now, all the configurations in $\{ R_0\}'$ are combinations of one in $\{ A_0 \}'$ and another in $\{ B_1 \}'$. Although the converse is not true, we  shall assume that $F_{\{ A_0 \}' } \, F_{\{ B_1 \}' } $ is dominated by the loop combinations that are in $\{ R_0\}'$, so that we can approximate
\[
F_{\{ R_0  \}'}  \approx  F_{\{ A_0 \}' } \, F_{\{ B_1 \}' }   \;.
\]
This assumption, together with Eq. \eqref{appFp}, gives 
\begin{eqnarray}
1 + F_{\{ B_0 \} } & \approx  & (1 + F_{\{ A_0 \}'})( 1 + F_{\{ B_1 \}'} ) \;. 
\end{eqnarray}  
We can proceed with a partition of $\{ B_1 \}'$ into three subsets: $\{ A_1\}$, given by loops that intersect $S(\mathcal{C})$ once; $\{ B_2 \}'$, where individual loops intersect $S(\mathcal{C})$ at least twice, and $\{ R_1\}'$, formed by combinations with single occupation. Applying a similar sequence of approximations, we have
\begin{eqnarray}
1 + F_{\{ B_0 \} } & \approx  &  (1 + F_{\{ A_0 \}'}) ( 1 + F_{\{ A_1 \}'} )  ( 1 + F_{\{ B_2 \}'} ) 
\label{ap-1}  \\
& \approx  &  (1 + F_{\{ A_0 \}'}) ( 1 + F_{\{ A_1 \}'} )  ( 1 + F_{\{ A_2 \}'} )  ( 1 + F_{\{ B_3 \}'} ) \;,
\end{eqnarray}
and so on. To represent the Wilson loop in Eq. \eqref{652-14}, each factor  must be computed at $\alpha_\mu$ (with linking numbers) and then divided by the factor at 
$\alpha_\mu \equiv 0$ (without linking numbers). Then, using Eq. \eqref{ap-1},  we have 
\begin{eqnarray}
\langle \mathfrak{z}({\cal C})  \rangle_{\rm latt} 
	\approx  \frac{(1 + 
	F_{\{ A_1 \}'}(\alpha_\mu) )}{ (1 + F_{\{ A_1 \}'}(0) )} \, \frac{(1 + 
	F_{\{ B_2 \}'}(\alpha_\mu) )}{(1 + F_{\{ B_2 \}'}(0) )}  \;.
\label{837-90}
\end{eqnarray} 
Next, the single self-avoiding loops in $\{ A_1 \}'$ can be partitioned into subsets 
  $\{ \mathbf{z}\}'$ labelled by the link  $\mathbf{z}$, $\mathbf{z}+ \hat{1}$  where the loop intersects $S(\mathcal{C})$ ($b_1(\mathbf{z}) = \pm 1$). Other configurations in $\{ A_1 \}'$ are combinations of $2, 3, \dots , {\rm Area }/a^2$ loops in different sets 
$\{ \mathbf{z}\}$. Then, we estimate, 
\begin{equation}
1 + F_{\{ A_1 \}'}(\alpha_\mu)  \approx \prod_{\mathbf{z}} (1 + F_{\{ \mathbf{z} \}'}(\alpha_\mu))  \makebox[.4in]{,}
F_{\{ \mathbf{z} \}'}(\alpha_\mu) = \xi_{\mathbf z} \,   \cos {\frac{2\pi }{N} } \makebox[.4in]{,} \xi_{\mathbf z}  =\sum_{\{ z \}' } e ^{-  \sum_{\mathbf{x}, \mu} 
 \frac{1}{2\beta_{\rm V}} \, |b_\mu (\mathbf{x})|  }   \;, 
\label{loops-fact}
\end{equation} 
where we have used that for every loop there is a similar one with reversed orientation. Disregarding 
$F_{\{ B_2 \}'}$, originated from single loops with multiple crossings, 
 and
approximating $\xi_{\mathbf{z} } $ as $\mathbf{z}$-independent quantities $\approx \xi$, we arrive at an area law, 
\begin{eqnarray}
\langle \mathfrak{z}({\cal C}) \rangle_{\rm latt}  & \approx &  \frac{ \Big( 1+ \xi   \, \cos {\frac{2\pi }{N}  } 
	\Big)^{M_A}   }{  \Big( 1+ \xi   \Big)^{M_A}  }  = e^{  -\sigma A } 
 \makebox[.3in]{,}  \sigma = - \frac{4 \eta }{ \beta_c^2} \, \ln   \left(  1-\frac{2\, \xi}{1+\xi } \,  \sin^2 \left( \frac{\pi}{N}\right) \right)  \;,
\label{res1}
\end{eqnarray}
where $\beta_c \approx 0.454$, $M_A$ is the number of links that cross $S({\cal C})$, which span an area $A = M_A \,a^2$, and we have 
 used Eq. (\ref{be}). Then, for $SU(2)$ we get,
\begin{equation}
\sigma_{\rm SU(2)} =  \frac{4 \eta }{ \beta_c^2} \, \ln   \left(  \frac{1+\xi}{1-\xi }  \right)  \;.
\end{equation} 
Finally we notice that the result obtained in eqs. (\ref{837-90})  and (\ref{loops-fact}) can be cast into the form
	 \begin{eqnarray}
\langle \mathfrak{z}({\cal C}) \rangle_{\rm latt}  & \approx & \prod_{\mathbf{z}} ((1-2p_{\mathbf{z}} )+ p_{\mathbf z} \,  e^{i\frac{2\pi }{N} } + p_{\mathbf z} \,  e^{-i\frac{2\pi }{N} } )  \makebox[.5in]{,} p_{\mathbf{z}} = \frac{\xi_{\mathbf{z}}/2}{1+ \xi_{\mathbf{z}} } \;,
\end{eqnarray}
where $p_{\mathbf{z}}$ and $1-2p_{\mathbf{z}}$ can be thought of as the probability for the link $\mathbf{z}$, $\mathbf{z}+1$ to take the value $e^{i\frac{2\pi}{N}}$ ($e^{-i\frac{2\pi}{N}}$), and 
$1$, respectively.

When $N\geq 3$, center vortex ensembles involve branching and the effective model in Eq. (\ref{zetaeff}) (based on a single field $V$) is not expected to be applicable. In this respect, for quarks in a representation with $N$-ality $k$, the center element generated by center vortices with linking number $1$ would be $ [z (1)]^k $. Then, if the average were performed along the same lines, with the same ensemble, the result would be,
\begin{eqnarray}
\langle \mathfrak{z}({\cal C}) \rangle_{\rm latt}  & \approx &  e^{  -\sigma_k A }  \makebox[.5in]{,}  \sigma_k = - \frac{4 \eta }{ \beta_c^2} \, \ln   \left(  1-\frac{2\, \xi}{1+\xi } \,  \sin^2 \left( \frac{ k\pi}{N}\right) \right)  \;,
\end{eqnarray}
and for large $N$, and finite $k$, the string tension ratios would approach a {\it squared} sine law,
\begin{eqnarray}
\frac{ \sigma_k}{\sigma_1}  \approx \frac{ \sin^2 \left( \frac{ k\pi}{N}\right) }{  \sin^2 \left( \frac{ \pi}{N}\right) }   \;.
\label{sine2}
\end{eqnarray}
However, this is not the expected behavior for Yang-Mills theory (see Ref. \cite{Bringoltz:2008nd} and references therein).

\section{Summary and Conclusions}\label{sectVI}
\label{discussion} 

In this paper we have investigated the effective field representation of an ensemble of closed center vortex loops in D = 3 Euclidean spacetime dimensions, as they emerge in the continuum limit after center projection of lattice gauge theory.  
To account for the properties of center vortices extracted on the lattice, the vortex loops were equipped with tension $(\mu)$, stiffness $(1/\kappa)$ and repulsive contact interactions. Such an ensemble of vortex loops, with definite orientation of the vortex flux, can be mapped onto an effective field theory of a complex scalar field, which has a global U(1) symmetry. 

We mainly focused on the evaluation of the expectation value of the Wilson loop, i.e. of center elements of the gauge group. For $\kappa, \mu > 0$ small vortex loops are favoured and a perimeter law is obtained as expected. In this case, the modes of the effective field theory are massive. For $\kappa > 0$ and $\mu < 0$, large center vortex loops are favoured and the corresponding effective field theory undergoes the spontaneous breaking of the U(1) symmetry. 
In this phase, the presence of Goldstone bosons introduces large quantum fluctuations. To deal with them, we formulated the problem on a lattice, using the XY model or its Villain approximation, which have been studied in the context of classical statistical mechanics and superfluids.   At high temperatures, the calculation is a perturbative one involving combinatorics.   However, to make contact with the continuum, it is important to consider the lattice models at the critical point, placed at $\beta_c \approx 0.454$, which has been accessed via Monte Carlo simulations. As this point is approached, the simulations show that more and more loops are generated. The phase transition is driven by single loops becoming infinitely long rather than by the proliferation of multiple smaller loops. In addition, the contribution of configurations with multiply occupied links is numerically irrelevant with respect to that originated from larger loops with single occupation due to the difference in entropy \cite{kleinert1989gauge}. These properties led to an area law as an extensive property. An interesting point is to clarify at which stage the minimum area appears in the calculation.   
In this respect, the consideration of loops that intersect the surface $S(\mathcal{C})$ only  once seems to work better for the minimum area surface. This is an important ingredient we  have used in the derivation. In addition, relevant effects can be incorporated by including  non-oriented center vortices with (correlated) magnetic monopoles. No matter how small this component is, it will break the $U(1)$ symmetry explicitly. In the Wilson loop calculation, the implied scale would enable a solitonic-like saddle point, localized on the minimum area surface, plus surface fluctuations that lead to the L\"uscher term. Then, while the center vortex loops (oriented or non-oriented) are essential to provide a confining linear potential with $N$-ality, the correlated monopoles could be relevant to endow this potential with string-like features.   

\section*{Acknowledgements}

L.E.O. would like to acknowledge the Conselho Nacional de Desenvolvimento Cient\'{\i}fico e Tecnol\'{o}gico (CNPq), CAPES, and FAPERJ for the financial support.


\end{document}